\documentclass[a4paper,11pt]{article}
\usepackage{jinstpub}
\usepackage{lineno}

\usepackage[utf8]{inputenc} 
\usepackage[T1]{fontenc}    
\usepackage{hyperref}       
\usepackage{url}            
\usepackage{booktabs}       
\usepackage{nicefrac}       
\usepackage{microtype}      
\usepackage{graphicx}       
\graphicspath{{figures/}}
\usepackage{bm}
\usepackage{makecell}
\usepackage{xspace}
\usepackage[inline]{enumitem}
\usepackage{paralist}
\usepackage{url}
\usepackage{hyperref}

\usepackage{siunitx}
\usepackage{xcolor}
\usepackage{tikz}
\usetikzlibrary{arrows.meta,positioning,fit}
\usepackage{tikz}
\usetikzlibrary{
    arrows.meta,
    positioning,
    calc,
    fit,
    decorations.pathreplacing
}

\newcommand{\vect}[1]{\boldsymbol{#1}}
\newcommand{\mum}{\ensuremath{\,\mu\text{m}}\xspace}
\newcommand{\nm}{\ensuremath{\,\text{nm}}\xspace}
\newcommand{\mm}{\ensuremath{\,\text{mm}}\xspace}

\newcommand{\volt}{\ensuremath{\,\text{V}}\xspace}

\newcommand{\ps}{\ensuremath{\,\text{ps}}\xspace}

\newcommand{\GSps}{\ensuremath{\,\text{GS/s}}\xspace}

\newcommand{\papertitle}{%
  GeoRead: Geometry-Aware Transformer for Continuous AC-LGAD Readout Reconstruction%
}

\title{\papertitle}

\author[1]{Lixing Wang,}
\author[2]{Leena Diehl,}
\author[1]{Spencer Ellis,}
\author[3]{Marcos Fern\'andez Garc\'ia,}
\author[4]{Gabriele Giacomini,}
\author[2]{Vagelis Gkougkousis,}
\author[2]{Ben Kilminster,}
\author[1]{Jingyu Luo,}
\author[2]{Anna Macchiolo,}
\author[1]{Yajna Mahabeer,}
\author[2]{Yevhenii Padniuk,}
\author[2]{Daria Senina,}
\author[1]{Samantha Sunnarborg,}
\author[1]{Jessica Tang,}
\author[4]{Alessandro Tricoli,}
\author[1]{Don C. Wong,}
\author[1,*]{Gaetano Barone\note[*]{Corresponding author.}}

\affiliation[1]{Brown University}
\affiliation[2]{University of Zurich}
\affiliation[3]{Instituto de F\'isica de Cantabria (CSIC-UC)}
\affiliation[4]{Brookhaven National Laboratory}

\abstract{
AC-coupled Low-Gain Avalanche Diodes (AC-LGADs) are promising silicon sensor technologies for high-precision four-dimensional tracking.
Position reconstruction in AC-LGADs relies on correlations among waveforms induced on multiple readout pads, where analytically solving the signal propagation becomes difficult, motivating data-based machine learning reconstruction. Existing machine learning approaches generally assume a fixed readout pad layout and predict one hit from an isolated, temporally aligned waveform window. These assumptions limit their flexibility across sensor designs and applicability to continuous readout where signal arrival times are not known in advance and multiple hits might overlap in the same waveforms.
We introduce GeoRead, a spatial-temporal Transformer enabling multi-event continuous position prediction and uncertainty estimation. We incorporate relative pad positions into attention and produces time-resolved hit position estimates with predictive uncertainties. Time-jitter and additive event-mixing augmentations are used to train the model on temporally misaligned and overlapping waveforms synthesized from single-event measurements.
Using four-channel transient-current-technique (TCT) data of a 500\mum-pitch AC-LGAD, GeoRead achieves median position errors of 5.8\mum within the convex hull of the readout-pad centers and 8.2\mum over the full scanned region at 190\volt. The model maintains comparable accuracy under temporal shifts and demonstrates time-resolved reconstruction of superposed events, without predefined per-event window. These results provide a basis for extending waveform-based position reconstruction toward continuous readout and flexible sensor geometries.
}

\begin{document}
\maketitle
\flushbottom

\section{Introduction}

Precision timing and spatial resolution are indispensable for next-generation experiments
such as the HL-LHC~\cite{CMS_TDR_MTD,CMSCollaboration_2008}, future $e^+e^-$ colliders,
and the EIC, where 4D tracking suppresses pile-up, enables time-of-flight particle
identification, and reconstructs dense vertices; similar needs arise in fixed-target
~\cite{ROYON2024168886} and space missions~\cite{GAUTIER2021165599,WU20192672}, and are
highlighted in community roadmaps~\cite{ECFA_Roadmap_2021}.

Low-Gain Avalanche Diodes (LGADs) and AC-coupled Low-Gain Avalanche Diodes (AC-LGADs)
are silicon sensors with internal gain that address these requirements. LGADs meet timing
requirements with $\mathcal{O}(20\text{--}50)$~\unit{ps} performance~\cite{Pellegrini2014,Wu_2020,
cartiglia2020,investigationchargesharing,lgad1,lgad2}, yet typical $\sim$$1\times1$$\mm^2$
pads limit spatial resolution. In the AC-LGAD variant~\cite{Mandurrino2020_NIMA959,
tornago2021_nima1003_165319,TORNAGO2021165319}, the AC pad is coupled to the resistive
$\mathrm{n}^+$ layer through a dielectric layer; the gain layer is not segmented, yielding
a 100\% fill factor that preserves LGAD timing performance while a resistive charge-sharing
mechanism distributes induced signals across multiple readout pads. This enables spatial
interpolation with a readout pitch larger than the desired position resolution. We target
high-energy physics applications in low-to-moderate radiation regimes, including FCC-ee
and upgrades to LHC experiments. For the High-Luminosity LHC, possible applications
include the LHCb VELO upgrade and a CMS tracker Phase-3 upgrade, where AC-LGADs could
extend timing capabilities in the forward region, increase rapidity coverage, or replace
one or two disks with timing-capable instrumentation. For FCC-ee, timing in the outermost
silicon layers could enhance particle identification and reduce the systematic uncertainty
on beam energy. Additional applications include time-of-flight systems and medical
detectors.

The correlations among pad signals depend on hit position, readout geometry, and signal
propagation through the resistive layer, making reconstruction particularly challenging.
Early machine-learning efforts using pulse height or time-over-threshold as features
improved robustness but still miss correlations between waveform shape, rise time, and
Landau fluctuations---undoubtedly, more information remains to be extracted, pushing toward
the noise limit. More recent work has demonstrated significant progress:
\begin{inparaenum}[(i)]  \item dense neural networks achieve 2--10\mum at 100--200\mum pitch ~\cite{siviero2021,investigationchargesharing},
  \item random forests reach $\sim$8\mum at 200\mum pitch~\cite{tornago2023},
  \item neural regression in beam tests delivers 5--10~$\mu$m
        ~\cite{heller2022_aclgad_bnl_hpk_jinst}, and
  \item previous work on full-waveform neural networks shows preliminary performance of
        $\sim10~\mu$m at $500\times500~\mu$m pitch~\cite{baronedrd3,Barone2025}.
\end{inparaenum}
Existing approaches, whether using handcrafted features~\cite{siviero2021,
siviero2022rsd16ch,tornago2023,investigationchargesharing} or full waveforms~\cite{
source_chinaxiv_202602_00195,aoki2026resistive}, offer a means of extracting spatial
information from correlated signals without explicitly inverting the charge-sharing
response.

However, most models use fixed-dimensional inputs associated with a prescribed set of
readout pads, implicitly encoding sensor geometry through channel identity. Changes in pad
number or arrangement therefore require modifications to the input representation and
potentially model retraining. Thus, these methods rely on hand-crafted fixed-length feature vectors, and they produce one position estimate per triggered event. Our previous topology-agnostic Transformer addressed this
limitation in part by incorporating absolute pad coordinates into the input~\cite{
aoki2026resistive}. A further limitation is the treatment of each recorded waveform window
as an isolated event, often aligned to the pulse peak, with one spatial prediction per
window. In continuous detector readout, hit arrival times are not necessarily known in
advance, and signals from successive particles may overlap. Event-level reconstruction
consequently depends on additional pulse identification and window selection. A
reconstruction model that directly processes temporally unaligned multi-channel signals
could reduce this dependence on preprocessing while preserving the spatial information
carried by the waveform evolution. More broadly other recent work has moved from hand-crafted features to raw, full-waveform modeling, such as the creation of 1D convolution neural network backbone followed by a fully connected head to predict two-dimensional hit positions from readout waveforms of a 300\mum-pitch AC-LGAD~\cite{source_chinaxiv_202602_00195}.  

In this work, we introduce GeoRead, a geometry-aware spatial-temporal Transformer for time-resolved AC-LGAD position
reconstruction. GeoRead is a flexible, high-throughput machine-learning-based readout. Building on our
previous topology-agnostic model, GeoRead incorporates pairwise relative pad geometry
directly into attention, following an approach related to the Particle Transformer~\cite{
qu2024particletransformerjettagging}. Multi-channel waveforms are divided into temporal
tokens, and causal attention with rotary position encoding~\cite{SU2024127063} enables
predictions from the signal history. The model produces both a hit-position estimate and
a predictive covariance for each time step, allowing native reconstruction of multi-event
waveforms. To relax the assumptions of peak alignment and isolated events, we train using
random temporal shifts and additive superposition of independent event waveforms; the
resulting architecture accepts explicit pad coordinates and supports temporally overlapping
signals without predefined event windows. As demonstrated by our prior algorithms~\cite{aoki2026resistive},  GeoRead the full waveform carries rich information not captured by binary readout or by a small number of amplitude-derived features.

We evaluate GeoRead using TCT measurements from a 500\mum-pitch AC-LGAD across bias
voltages from 100 to 190\volt.  At the architectural level, GeoRead builds on the Transformer~\cite{vaswani2017attention}, uses rotary position encoding to represent temporal order~\cite{SU2024127063}, and injects pairwise pad geometry into attention in a way conceptually related to the Particle Transformer~\cite{qu2024particletransformerjettagging}. We characterize spatial reconstruction performance within
and outside the convex hull of the readout-pad centers, investigate robustness to temporal
misalignment, and demonstrate continuous predictions on synthetically mixed multi-event
waveforms. We further examine the relationship between predicted uncertainty and
reconstruction error, and compare the model with its full-waveform variant as baseline.
Additional results on test-beam measurements provide an assessment under particle-beam
conditions. Together, these studies examine whether time-resolved reconstruction and
explicit geometry conditioning can extend the capabilities of full-waveform methods while
retaining micrometer-scale spatial precision.

In summary, our main contributions are: (i) a geometry-aware formulation that decouples absolute sensor geometry from reconstruction task; (ii) a spatial-temporal formulation for native multi-event continuous AC-LGAD position reconstruction; (iii) a training
pipeline based on time jitter and event mixing that relaxes peak-alignment and single-event
assumptions; and (iv) predictive uncertainty that reflects reconstruction quality.


%
%
%

\section{Dataset and Problem Formulation}
\label{sec:data}

\subsection{Dataset from Transient Current Techniques}

We collected the transient-current-technique (TCT) dataset using a Particulars scanning-TCT system and an Broohkaven National Laboratory-fabricated AC-LGAD with a pitch of 500\mum $\times$ 500\mum and pad dimensions of $200\mum \times 200\mum$. The sensor was scanned using an infrared laser with a wavelength of 1064\nm over a full region of interest (ROI) of $1\mm \times 1\mm$, with a grid step size of 20\mum. The laser spot had an approximately $9\mum$ Gaussian minimum beam waist, and the laser intensity was adjusted to produce a signal corresponding to approximately 33 minimum-ionizing particles (MIPs). Four AC-LGAD readout channels, whose pad centers form a square convex hull, were read out using a Chubut 2 board with two amplification stages per channel. Waveforms were digitized using a LeCroy 640Zi oscilloscope at 20\GSps, corresponding to a sampling interval of 50\ps. 

Each event contains four time-series waveforms, one per channel, together with a ground-truth two-dimensional hit coordinate. Each waveform is corrected for DC offset using a segment of instrumental noise at its beginning. Each event's peak is aligned in time. The waveforms were collected across detector bias voltages of 100, 120, 130, 140, 160, 170, 180, and 190\volt. Each voltage contains approximately 50K events, and each is partitioned to training, validation, and test set with 60/20/20 ratio. Each truth coordinate is shifted such that the bottom-left corner is the origin, and the bias voltages are normalized to be between -1 and 1. The sensor's physical pad coordinates are automatically computed from train split event amplitude distribution of each channel. 

To simulate misaligned, multi-event waveforms, we employ two data augmentations which break the assumption that each event is aligned in time and each window contains exactly one event:

\paragraph{Time Jitter} We simulate misaligned events by choosing a random window around each event's peak. We first select a parent window to save memory, covering 40 samples before and 128 samples after the event peak. Then, a smaller window of 128 samples is selected by randomly picking a start from 0th to 18th sample point of the parent window. In evaluations without time jitter, we fix the start at 9th sample.

\paragraph{Event Mixing} We construct multi-event waveforms by superimposing temporally shifted event waveforms. For a mixture of $N_{\text{evt}}$ events, each event is first extracted over an extended window of $128 \cdot N_{\text{evt}}$ samples, where $128$ is the single-event window length. The first event starts at zero, while subsequent events are added at sampled temporal offsets. If time jitter is enabled, then the offsets are randomized as well. This additive mixing simulates pile-up from close events with overlapping signals. 

We also collected a test-beam dataset at the CERN SPS using 120~\unit{GeV}/$c$ pions with reference hit positions provided by a MIMOSA-26 telescope, and applied our model framework to these data to assess its capabilities in dirtier, more realistic settings. The rest of this study will mainly focus on TCT measurements, which provide a cleaner and more controlled setting for probing the GeoRead architecture's capabilities. 

\subsection{Definition of the Spatial-temporal Reconstruction Task}

Let $x_{c,t}$ denote the waveform sample from readout channel $c$ at sample time $t$, and let $\vect{r}=(x,y)$ denote the hit coordinate. The waveform is divided into temporal chunks of $K$ samples. For a four-channel input of length $T$, this produces $4T/K$ channel-time tokens. The network returns a sequence
\begin{equation}
    \left\{\hat{\vect{r}}_m,\hat{\Sigma}_m\right\}_{m=1}^{T/K},
\end{equation}
where $m$ indexes temporal chunks, $\hat{\vect{r}}_m$ is a point estimate, and $\hat{\Sigma}_m$ is a covariance prediction. For single-event sequences, the target position remains constant throughout the waveform. For synthetically mixed sequences, the target of each time step is assigned to ground truth position of the last event seen so far, and the model might use signal history as context to resolve superposed waveforms.

\section{Model Architecture}
\label{sec:model}
\begin{figure}[t]
    \centering
    \includegraphics[width=0.9\textwidth]{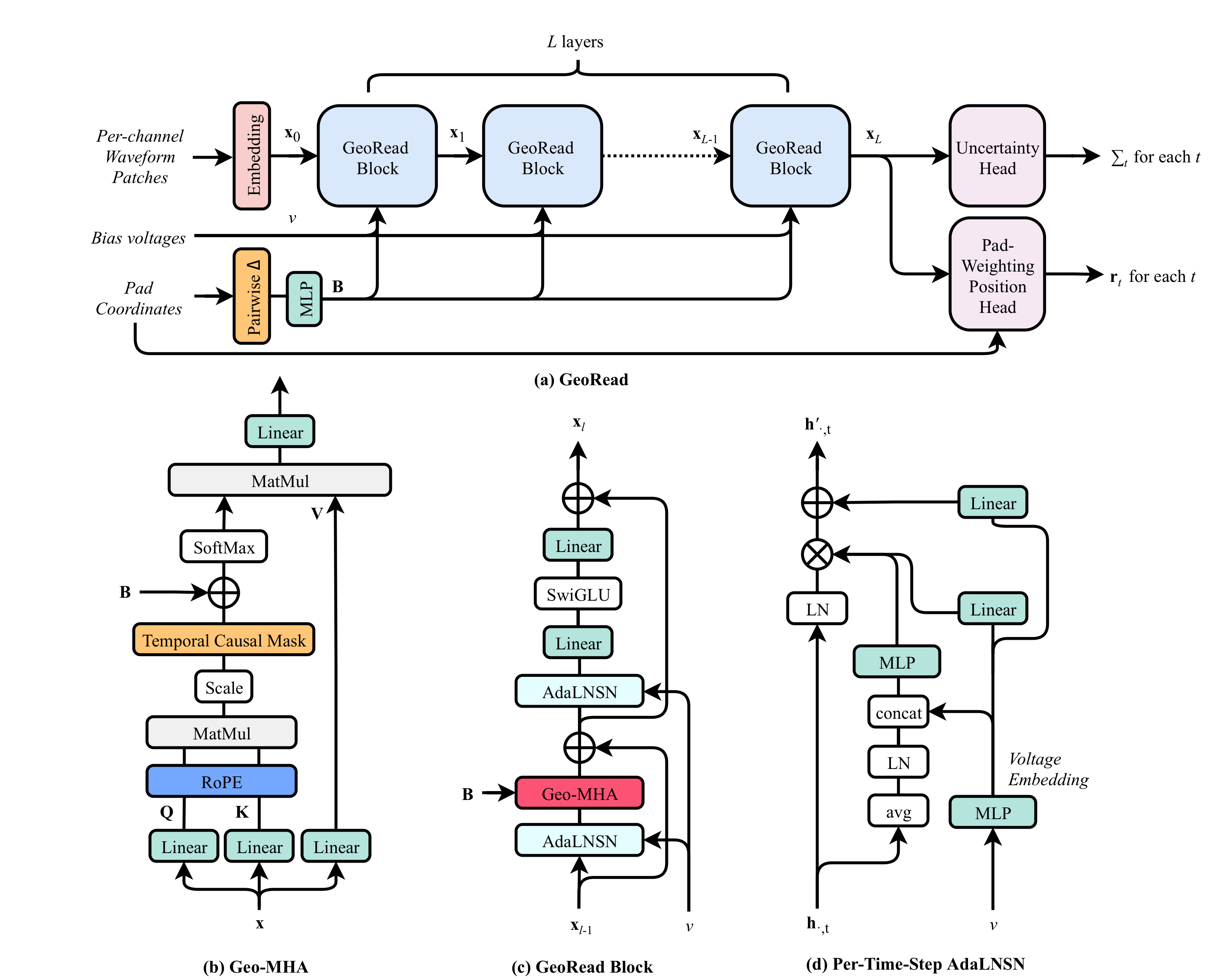}
    \caption{
        \textbf{Overview of the GeoRead architecture.}\\[0.25em]
        \textbf{(a)} Each readout channel is divided into non-overlapping temporal waveform patches and projected into a sequence of tokens. Physical pad coordinates are converted into pairwise coordinate differences and mapped by an MLP to attention biases shared across the $L$ GeoRead blocks. Bias voltage is provided as an additional conditioning variable. The final hidden states are processed by an uncertainty head and a pad-weighting position head, with the latter explicitly using the physical pad coordinates to construct the spatial prediction.\\[0.25em]
        \textbf{(b)} Geometry-aware multi-head attention (Geo-MHA). Rotary position encoding (RoPE) is applied to the queries and keys, while a temporal causal mask prevents attention to future time indices. The learned attention bias is added to the logits before softmax.\\[0.25em]
        \textbf{(c)} Each GeoRead block follows a pre-normalization Transformer structure with per-time-step AdaLNSN, Geo-MHA, residual connections, and a SwiGLU feed-forward network.\\[0.25em]
        \textbf{(d)} Per-time-step AdaLNSN. For each time index $t$, hidden states are aggregated across pads only at the same time index. Their mean representation is combined with an embedding of the bias voltage to determine the sample-dependent modulation, in addition to the voltage-only affine transformation.
    }
    \label{fig:architecture}
\end{figure}

An overview of the GeoRead architecture is shown in \autoref{fig:architecture}. GeoRead combines temporal waveform tokenization, geometry-aware attention, voltage-conditioned normalization, and per-time-step position and uncertainty heads. We describe each component in detail below.

\paragraph{Input and Position Encoding}

The model takes a multi-channel time-series waveform, sensor's bias voltage, and each channel's physical coordinates as inputs. For every channel, consecutive $K$ waveform samples are projected to the Transformer hidden dimension. Tokens from the same temporal chunk share a time index. RoPE is applied using this shared time index, so multiple channel tokens at a given temporal location receive the same temporal phase. 

\paragraph{Spatial Attention Bias}

We explicitly inject the sensor geometry in two ways: relative pad geometry via attention bias, and absolute pad coordinates via the position head.

The readout channels have known pad coordinates $\vect{r}_c$. We introduce geometry through the attention bias assuming symmetry between channels' response property. For a pair of channels $(c_i,c_j)$, a small multilayer perceptron maps their coordinate displacement to a head-dependent bias $b_h(c_i,c_j)$ shared across all layers. For attention head $h$,
\begin{equation}
    A_{ij}^{(h)} = \frac{\vect q_i^{(h)}\cdot \vect k_j^{(h)}}{\sqrt{d_h}} + b_h(c_i,c_j) + M_{ij},
\end{equation}
where $M_{ij}$ is the causal mask. 

\paragraph{Pad-weighting Position Head and Uncertainty Head}

The position head outputs a point prediction for each time index. At each time index $m$, the hidden state for channel $c$ is projected to a scalar score $w_{m,c}$. 
The predicted position is a weighted average of known pad coordinates,
\begin{equation}
    \hat{\vect r}_m = \frac{\sum_c w_{m,c}\vect r_c}{\sum_c w_{m,c}}.
\end{equation}
where $\vect r_c$ denotes the pad coordinate for channel $c$, and $\hat{\vect r}_m$ denotes the point prediction at time index $m$.
Note that the formula enables out-of-hull prediction by allowing negative $w_{m,c}$. The position head has no trainable parameters other than the projection layer.

A separate uncertainty head acts on the average final hidden state at the same time index and predicts a positive-definite $2\times2$ covariance matrix $\hat{\Sigma}_m$ through a Cholesky parameterization. The model is trained using the two-dimensional Gaussian negative log-likelihood,
\begin{equation}
    \mathcal{L}_{\mathrm{NLL}}
    = \frac{1}{M}\sum_{m=1}^{M}
    \frac{1}{2}
    \left[
        (\vect r_m-\hat{\vect r}_m)^\top
        \hat{\Sigma}_m^{-1}
        (\vect r_m-\hat{\vect r}_m)
        + \log\det\hat{\Sigma}_m
    \right],
\end{equation}
where $M$ denotes the total number of time indices, $\vect r_m$ indicates ground truth hit coordinate, and $\hat{\Sigma}_m^{-1}$ denotes the inverse of the predicted covariance matrix at time index $m$.
For single-event samples, $\vect r_m$ is constant in time; for mixed samples it is the ground truth position of the last event seen so far.

\paragraph{Temporal Causal Masking}

At time index $m$, tokens are allowed to attend to tokens from the same or earlier time indices, but not to future ones, and thus $\hat{\vect r}_m$ can be interpreted as a prediction available after observing the waveform only up to time $m$. Tokens within the same temporal chunk remain mutually visible.

\paragraph{Adaptive Normalization}

The model uses AdaLNSN (Adaptive Layer Normalization and Sample Normalization), an adaptive normalization variant developed for AC-LGAD reconstruction. The normalization parameters depend on detector bias voltage, while an additional signal-state modulation is mapped from the average hidden states at respective time indices. This could allow the network to respond to signal-to-noise ratio changes coming from random charge fluctuations, which are independent from bias voltage.

\section{Results and Performance}

\paragraph{Implementation Details}
\label{sec:training}

We use 8 Transformer layers, hidden dimension 16, 4 attention heads, dropout 0.1, time-token size 8, batch size 512, and AdamW with a cosine annealing learning rate starting from $5\times10^{-4}$. Models are trained for 100 epochs. No normalization or amplitude scaling is applied to the waveforms. We train the model jointly on all available voltage points and respectively evaluate on each voltage.
The model consists of 67,096 parameters, which is very parameter efficient and can fit in compact devices for inference.

\label{sec:results}
\paragraph{Continuous Multi-event Prediction}

\begin{figure}[htbp]
    \centering
    \includegraphics[width=\textwidth]{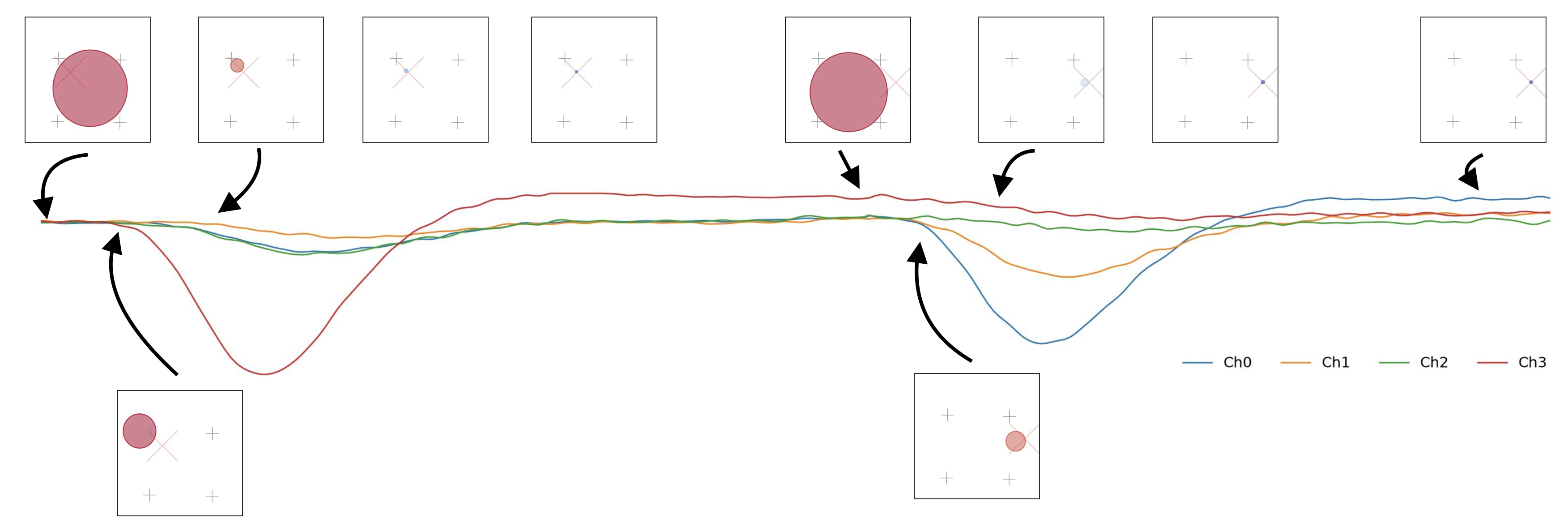}
    \caption{\textbf{Spatial-temporal predictions on a mixed two-event sequence example.}
    The square plots represent the scanned region of the sensor. The plus signs show pad centers, and the cross shows ground truth position. Each output of GeoRead is illustrated as an ellipse, whose center shows predicted position and size shows uncertainty. The color of ellipse shows reconstruction error.}
    \label{fig:multievent}
\end{figure}

We conduct qualitative evaluation on multi-event samples. As shown in \autoref{fig:multievent}, on superposed two-event examples, the model maintains a stable estimate across time and updates its prediction after the second event becomes observable. The transition is accompanied by a temporary increase in predicted uncertainty, and the uncertainty drops once the peak is seen. More examples are included in appendix.

\paragraph{Single-event Spatial Resolution}

\begin{figure}[t]
    \centering
    \includegraphics[width=\textwidth]{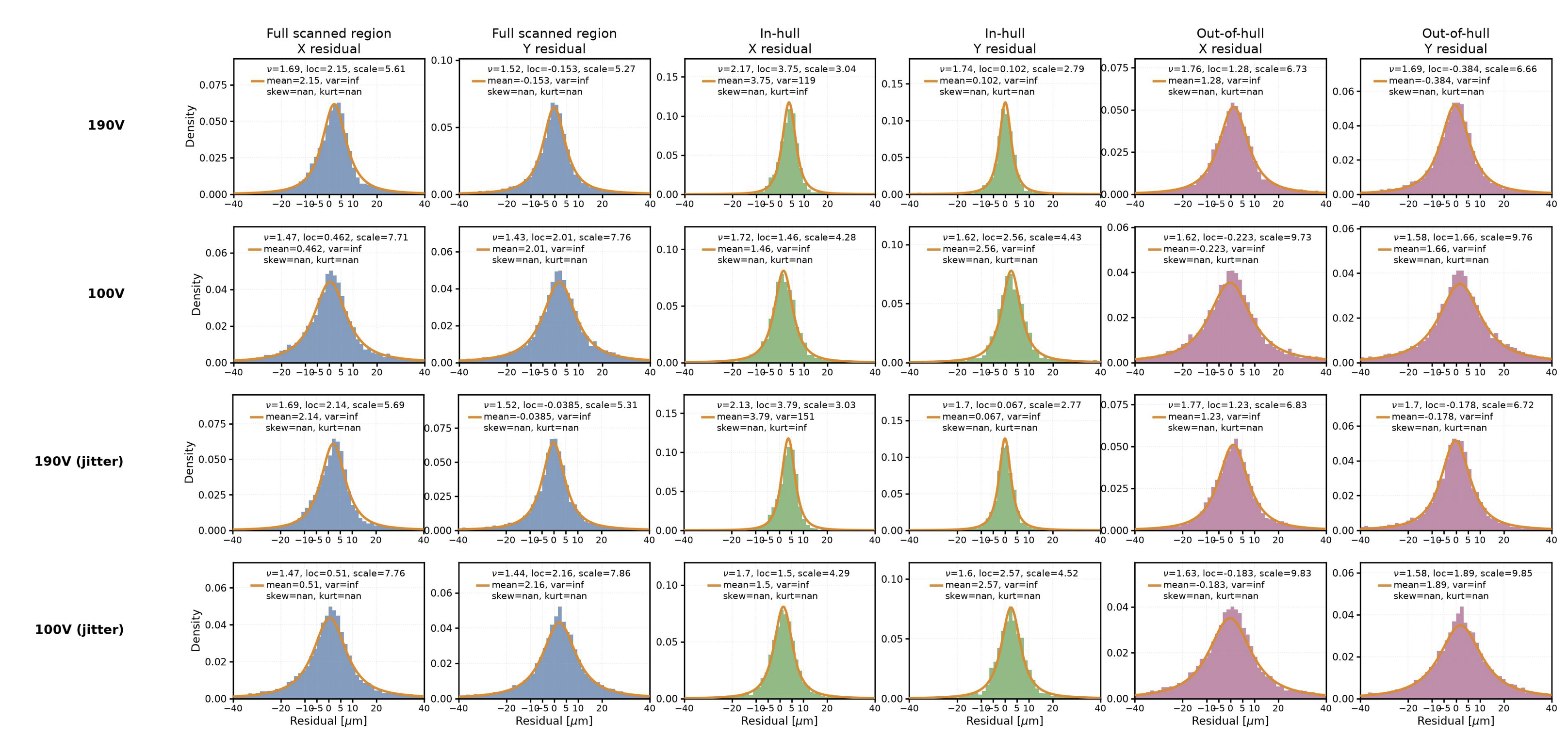}
    \caption{
        \textbf{Student-t fits summary for distribution of reconstruction residual in x and y directions at 190 and 100\volt for full scanned region, in-hull region, and out-of-hull region.} 
        The x-axis represents residual of model prediction in \mum, and the y-axis represents probability density. The test waveforms here are single-event ones. Full results are attached in appendix.
    }
    \label{fig:student-t-summary}
\end{figure}

We conduct quantitative evaluation on the model's reconstruction accuracy on single-event waveforms. The model's prediction for such an event is chosen to be the one with smallest scalar uncertainty $\sqrt{\operatorname{tr}{\Sigma}}$. We report percentiles of reconstruction error distribution on full scanned region, in-hull region, and out-of-hull region. The in-hull region corresponds to the region inside the convex hull formed by the pad centers, where charge-sharing information is expected to be most abundant. \autoref{tab:finalcpd} summarizes the performance over time-jittered single-event waveforms as well as peak-aligned single-event waveforms, and the stuent-t analysis of residual distribution is summarized in \autoref{fig:student-t-summary}. The model achieves 5.8\mum of median reconstruction error in in-hull region, and 8.2\mum over full scanned region. Time-jitter evaluation shows comparable reconstruction accuracy. The performance improves with higher bias voltages, and in-hull reconstruction outperforms out-of-hull reconstruction. More results are provided in appendix.

\newcommand{\jit}[1]{\textcolor{gray!65}{\small \textit{(#1)}}}
\newcommand{\tail}[1]{\textcolor{gray!75}{#1}}
    
\begin{table}[t]
    \centering
    \caption{
    Reconstruction-error 50th and 99th percentiles in the full scanned region, in-hull region, and out-of-hull region. All values are in \mum. Values in parentheses correspond to time-jitter evaluation.
    }
    \label{tab:finalcpd}
    \setlength{\tabcolsep}{3pt}

    \begin{tabular}{c cc cc cc}
        \toprule
        &
        \multicolumn{2}{c}{Full ROI} &
        \multicolumn{2}{c}{In-hull} &
        \multicolumn{2}{c}{Out-of-hull} \\
        \cmidrule(lr){2-3}
        \cmidrule(lr){4-5}
        \cmidrule(lr){6-7}

        Voltage [\volt]
        & $r_{50}$ & \tail{$r_{99}$}
        & $r_{50}$ & \tail{$r_{99}$}
        & $r_{50}$ & \tail{$r_{99}$} \\
        \midrule

        190
        & 8.2 \jit{8.3} & \tail{103.1 \jit{104.9}}
        & 5.8 \jit{5.7} & \tail{51.7 \jit{52.7}}
        & 9.5 \jit{9.8} & \tail{114.0 \jit{117.3}} \\

        180
        & 8.6 \jit{8.5} & \tail{107.8 \jit{105.4}}
        & 5.9 \jit{5.9} & \tail{55.0 \jit{55.4}}
        & 10.1 \jit{9.9} & \tail{115.9 \jit{115.7}} \\

        170
        & 8.8 \jit{8.9} & \tail{104.7 \jit{106.0}}
        & 6.0 \jit{5.9} & \tail{54.3 \jit{54.8}}
        & 10.5 \jit{10.5} & \tail{115.0 \jit{119.1}} \\

        160
        & 8.8 \jit{8.8} & \tail{101.9 \jit{101.8}}
        & 5.9 \jit{5.8} & \tail{62.3 \jit{60.0}}
        & 10.4 \jit{10.5} & \tail{118.2 \jit{121.6}} \\

        140
        & 9.7 \jit{9.8} & \tail{109.4 \jit{110.0}}
        & 6.1 \jit{6.1} & \tail{58.5 \jit{59.1}}
        & 11.9 \jit{11.9} & \tail{124.0 \jit{125.1}} \\

        130
        & 11.0 \jit{11.0} & \tail{136.5 \jit{144.9}}
        & 7.1 \jit{7.1} & \tail{70.8 \jit{68.7}}
        & 13.1 \jit{13.0} & \tail{149.7 \jit{158.3}} \\

        120
        & 10.7 \jit{10.9} & \tail{124.6 \jit{121.5}}
        & 6.4 \jit{6.5} & \tail{63.2 \jit{63.2}}
        & 13.1 \jit{13.3} & \tail{138.0 \jit{131.0}} \\

        100
        & 12.0 \jit{12.2} & \tail{160.3 \jit{163.1}}
        & 7.3 \jit{7.4} & \tail{98.9 \jit{99.6}}
        & 14.5 \jit{14.7} & \tail{165.6 \jit{170.2}} \\

        \bottomrule
    \end{tabular}
\end{table}

\paragraph{Temporal Evolution of Reconstruction Quality}

\begin{figure}[t]
    \centering
    \includegraphics[width=0.95\textwidth]{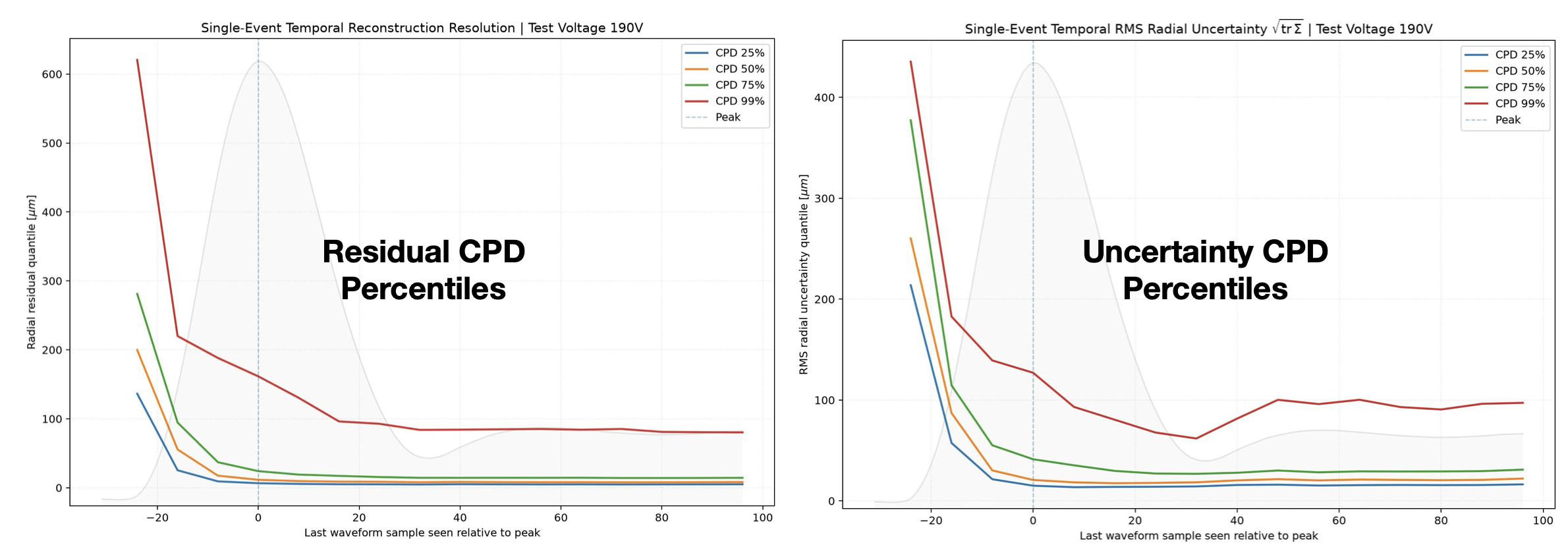}
    \caption{\textbf{Temporal development of reconstruction error and predicted uncertainty on aligned single-event samples.} The x-axis represents time, and the y-axis represents the scalar uncertainty $\sqrt{\operatorname{tr}{\Sigma}}$ or error in \mum. The blue, orange, green, and red curves represent 25th, 50th, 75th and 99th percentile of the reconstruction-error or uncertainty distribution at each time step, respectively. The gray curve shows the average waveform shape. The temporal pattern of predicted uncertainty matched closely with that of the actual error, while the model reports larger uncertainties as the waveform decays.}
    \label{fig:temporal}
\end{figure}

We qualitatively examine the temporal evolution of predicted uncertainty and actual reconstruction error. As shown in \autoref{fig:temporal}, the distribution of predicted uncertainty closely follows that of the reconstruction error. They are initially large, then decrease sharply and stabilize around the signal peak. Interestingly, as the waveform decays after the peak, the predicted uncertainty increases while the reconstruction error remains low. This behavior may reflect the model's uncertainty about the arrival of subsequent events or increasing ambiguity in the decaying waveform.

\paragraph{Correlation between Reconstruction Error and Uncertainty}

\begin{figure}[t]
    \centering 
    \includegraphics[width=0.9\textwidth]{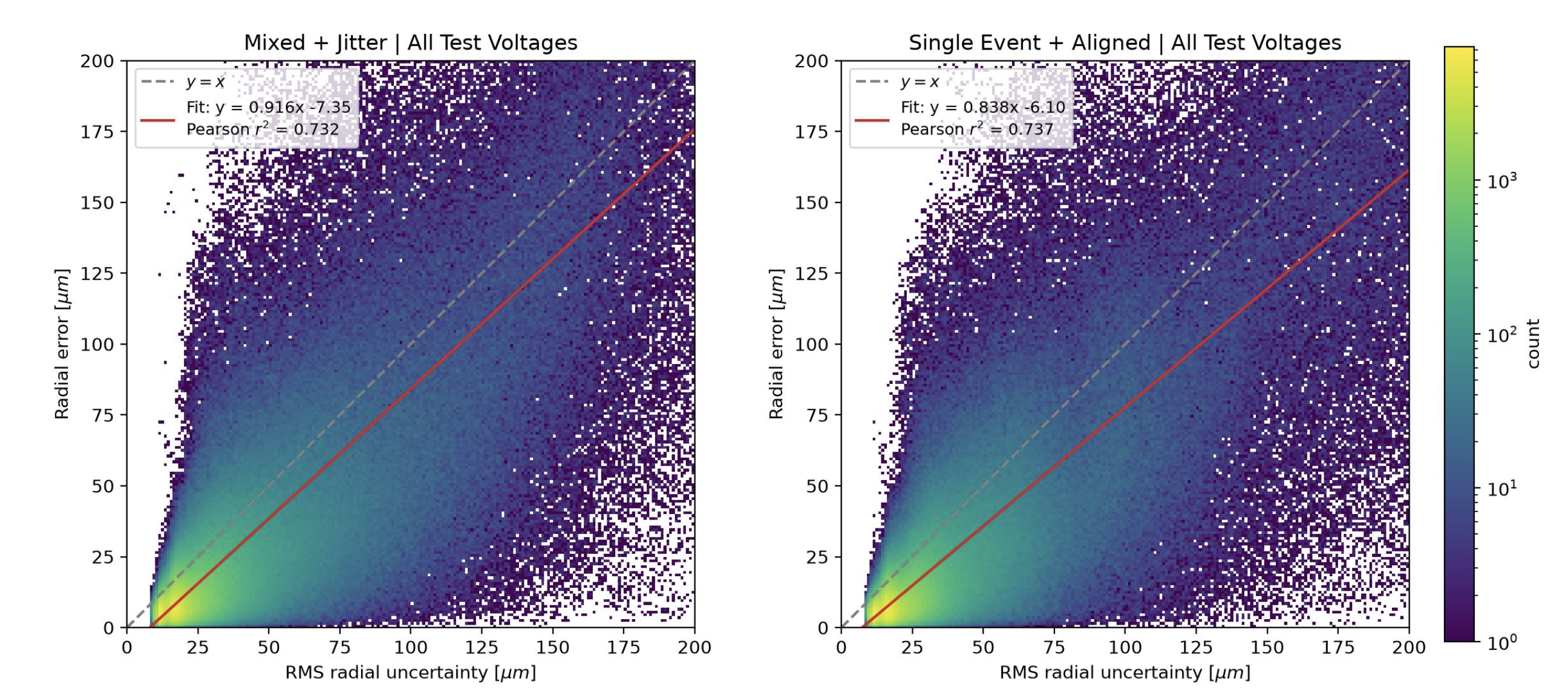}
    \caption{\textbf{Relationship between reconstruction error and predicted uncertainty on time jittered multi-event samples and on aligned single-event samples.} The x-axis shows the scalar uncertainty $\sqrt{\operatorname{tr}{\Sigma}}$, the y-axis shows the radial reconstruction error, both in \mum. Color indicates the number of predictions per bin on a logarithmic scale. The squared Pearson correlation coefficients ($r^2$) are 0.732 and 0.737, respectively.}
    \label{fig:error_uncertainty}
\end{figure}

We further evaluate whether the predicted uncertainty reflects the actual reconstruction quality on both temporally jittered multi-event samples and peak-aligned single-event samples. As shown in \autoref{fig:error_uncertainty}, the scalar uncertainty $\sqrt{\operatorname{tr}{\Sigma}}$ exhibits a strong positive correlation with the radial reconstruction error, with squared Pearson correlation coefficients ($r^2$) of 0.732 and 0.737, respectively, suggesting that the uncertainty estimates remain informative of reconstruction quality. 

\section{Ablation Studies}
\label{sec:ablations}

\paragraph{Effect of Time Jitter and Event Mixing}

\autoref{tab:augablation} separates the effects of the two training augmentations. Even without train-time augmentation, GeoRead exhibits intrinsic robustness against time jitter. Event mixing only and time jitter only both add a cost, where the model potentially needs more capacity to identify and ignore irrelevant context and infer where the informative pulse lies. Yet, training with both augmentations performs better than using either one, for gaining both the new multi-event and misalignment capabilities.

\begin{table}[t]
    \centering
    \caption{
    Data-augmentation ablation for GeoRead.
    Reported values are median reconstruction errors, in \mum.
    Values in parentheses correspond to time-jitter evaluation.
    }
    \label{tab:augablation}

    \begin{tabular}{l cc cc}
        \toprule
        &
        \multicolumn{2}{c}{$r_{50}$ In-hull} &
        \multicolumn{2}{c}{$r_{50}$ Full ROI} \\
        \cmidrule(lr){2-3}
        \cmidrule(lr){4-5}

        Training augmentation
        & 190\volt
        & 100\volt
        & 190\volt
        & 100\volt \\
        \midrule

        None
        & 4.2 \jit{5.4}
        & 5.8 \jit{6.4}
        & 6.3 \jit{7.7}
        & 9.7 \jit{10.7} \\

        Event mix only
        & 6.5 \jit{7.3}
        & 8.4 \jit{9.2}
        & 8.8 \jit{8.8}
        & 12.9 \jit{12.9} \\

        Time jitter only
        & 6.9 \jit{6.9}
        & 7.7 \jit{7.7}
        & 10.0 \jit{9.9}
        & 13.4 \jit{13.7} \\

        Event mix + time jitter
        & 5.8 \jit{5.7}
        & 7.3 \jit{7.4}
        & 8.2 \jit{8.3}
        & 12.0 \jit{12.2} \\

        \bottomrule
    \end{tabular}
\end{table}

\paragraph{Intrinsic Robustness to Temporal Misalignment}

We are interested in how the spatial-temporal version's performance compares with the full-waveform one. The full-waveform model is equivalent to setting the time-token size of GeoRead to the entire window size.
It receives one 128-sample waveform per channel. Each complete channel waveform is projected to one channel token, producing four tokens in total for the AC-LGAD involved. 

As shown in \autoref{tab:jitterrobustness}, the spatial-temporal GeoRead trained without time jitter degrades only modestly under time-jitter evaluation. Yet, the full-waveform model trained without jitter is sensitive to temporal shifts. Adding jitter augmentation to the full-waveform baseline largely restores its robustness. This suggests that short time tokens provide an architectural bias toward temporal shift robustness, while explicit jitter augmentation can supply a similar invariance to a full-waveform model. Note that the full-waveform model is slightly better in reconstruction quality, which could imply that temporal modeling is a more challenging task.

\begin{table}[t]
    \centering
    \caption{
    Robustness to temporal misalignment.
    Reported values are median reconstruction errors, in \mum.
    Values in parentheses correspond to time-jitter evaluation.
    }
    \label{tab:jitterrobustness}

    \begin{tabular}{l cc cc}
        \toprule
        &
        \multicolumn{2}{c}{$r_{50}$ In-hull} &
        \multicolumn{2}{c}{$r_{50}$ Full ROI} \\
        \cmidrule(lr){2-3}
        \cmidrule(lr){4-5}

        Model and training
        & 190\volt
        & 100\volt
        & 190\volt
        & 100\volt \\
        \midrule

        GeoRead, no augmentation
        & 4.2 \jit{5.4}
        & 5.8 \jit{6.4}
        & 6.3 \jit{7.7}
        & 9.7 \jit{10.7} \\

        Full-waveform, no augmentation
        & 3.3 \jit{4.9}
        & 5.3 \jit{6.5}
        & 5.3 \jit{9.5}
        & 9.1 \jit{11.9} \\

        Full-waveform + time jitter
        & 3.3 \jit{3.4}
        & 5.3 \jit{5.4}
        & 5.3 \jit{5.3}
        & 9.0 \jit{9.2} \\

        \bottomrule
    \end{tabular}
\end{table}

\paragraph{Sampling Rate Studies}
To investigate the effect of sampling rate on reconstruction performance, we uniformly downsample the input waveforms by factors of 2, 4, 8, and 16 after applying data augmentations, and train GeoRead for each from scratch. As shown in \autoref{tab:rasterization}, reconstruction accuracy remains broadly comparable to the original 20 GS/s baseline at sampling rates down to 2.5 GS/s, although some degradation is observed, particularly at 100\volt. Further reducing the sampling rate to 1.25 GS/s leads to a substantial increase in reconstruction error. These results suggest that the temporal sampling requirements of GeoRead may be relaxed considerably without a major loss of precision.
\begin{table}[t]
    \centering
    \caption{
    Performance at different sampling rates.
    Reported values are median reconstruction errors, in \mum.
    Values in parentheses correspond to time-jitter evaluation.
    }
    \label{tab:rasterization}

    \begin{tabular}{l cc cc}
        \toprule
        &
        \multicolumn{2}{c}{$r_{50}$ In-hull} &
        \multicolumn{2}{c}{$r_{50}$ Full ROI} \\
        \cmidrule(lr){2-3}
        \cmidrule(lr){4-5}

        Sampling rate
        & 190\volt
        & 100\volt
        & 190\volt
        & 100\volt \\
        \midrule

        1:1 (20 GS/s)
        & 5.8 \jit{5.7}
        & 7.3 \jit{7.4}
        & 8.2 \jit{8.3}
        & 12.0 \jit{12.2} \\

        1:2 (10 GS/s)
        & 6.9 \jit{7.1}
        & 9.0 \jit{9.0}
        & 10.7 \jit{10.5}
        & 15.2 \jit{15.4} \\

        1:4 (5 GS/s)
        & 6.1 \jit{6.2}
        & 8.4 \jit{7.9}
        & 8.8 \jit{9.0}
        & 13.5 \jit{13.8} \\

        1:8 (2.5 GS/s)
        & 5.8 \jit{6.1}
        & 8.7 \jit{8.6}
        & 8.5 \jit{8.8}
        & 16.0 \jit{15.9} \\

        1:16 (1.25 GS/s)
        & 10.6 \jit{11.6}
        & 12.1 \jit{12.4}
        & 22.2 \jit{24.0}
        & 26.0 \jit{26.3} \\

        \bottomrule
    \end{tabular}
\end{table}

\section{Conclusion}
\label{sec:discussion}

We presented GeoRead, a geometry-aware spatial-temporal Transformer for continuous AC-LGAD position reconstruction. The model incorporates relative pad geometry into attention and produces time-resolved position estimates with predictive uncertainties. Using TCT measurements from a 500\mum-pitch sensor, GeoRead achieves median position errors of 5.8\mum within the convex hull of the readout-pad centers and 8.2\mum over the full scanned region at 190\volt (\autoref{tab:finalcpd}). These results demonstrate that time-resolved reconstruction can retain micrometer-scale spatial precision while relaxing the fixed-window assumptions of earlier full-waveform approaches.

Temporal modeling introduces a trade-off in reconstruction accuracy. At 190\volt and over full scanned region, the unaugmented GeoRead achieves a median error of 6.3\mum on aligned inputs, compared with 5.3\mum for the full-waveform baseline. Nevertheless, GeoRead shows greater intrinsic robustness to temporal shifts, and training with both time jitter and additive event mixing yields a median error of 8.3\mum on jittered inputs (\autoref{tab:jitterrobustness}, \autoref{tab:augablation}). The model also demonstrates continuous position updates on synthetic multi-event waveforms formed by superimposing temporally shifted TCT signals (\autoref{fig:multievent}). These results support the feasibility of reconstructing successive hits without predefined event windows, with a moderate loss in single-event spatial accuracy.

The predicted uncertainty follows the temporal evolution of reconstruction error, increasing when the available signal is weak and decreasing as the waveform develops (\autoref{fig:temporal}). This information may be useful for assessing the reliability of individual position estimates in downstream reconstruction. The explicit geometry conditioning also allows the architecture to represent different readout layouts, although generalization across sensor geometries remains to be demonstrated. In addition to the TCT studies, application of the model architecture to CERN SPS test-beam data collected with 120 GeV/$c$ pions yields a median two-dimensional residual of 45.9\mum at 160\volt, compared with 134.5\mum for matrix inversion over a $1.0\times1.0$~mm$^2$ fiducial region. 

Several limitations remain. The multi-event evaluation relies on additive superposition of independently measured TCT waveforms. Although this produces overlapping signals, it assumes linear waveform superposition and does not capture all effects of pile-up in realistic detector operation. Validation with experimentally recorded overlapping particle signals is therefore still required. Furthermore, geometry-aware attention does not explicitly account for channel-dependent response variations, which may require additional calibration or conditioning. The computational cost and inference latency of continuous reconstruction also remain to be evaluated. Addressing these limitations will be important for assessing the use of GeoRead in continuous readout systems.

\section*{Acknowledgments}
We would like to thank the Solid State Detectors Lab of the EP-DT group at CERN for the support. This work was performed in part within the framework of the DRD3 Collaboration and was supported in part by the Brown University Undergraduate Teaching and Research Award (UTRA), Brown Research Seed Award No. GR300352, and the BEACON 2026 Brown Data Science Collaborative Award. This material is based upon work supported by the U.S. Department of Energy under grants DE-447 SC0012704, DE-SC443363, DE-SC426496 and DE-448 SC0020255. This research used resources of the Center for Functional Nanomaterials, which is a U.S. DOE Office of Science Facility, at Brookhaven National Laboratory under Contract No. DE-SC0012704.

\appendix

\section{Amplitude Distribution over Scanned Region}
\begin{figure}[p]
    \centering
    \includegraphics[width=\textwidth]{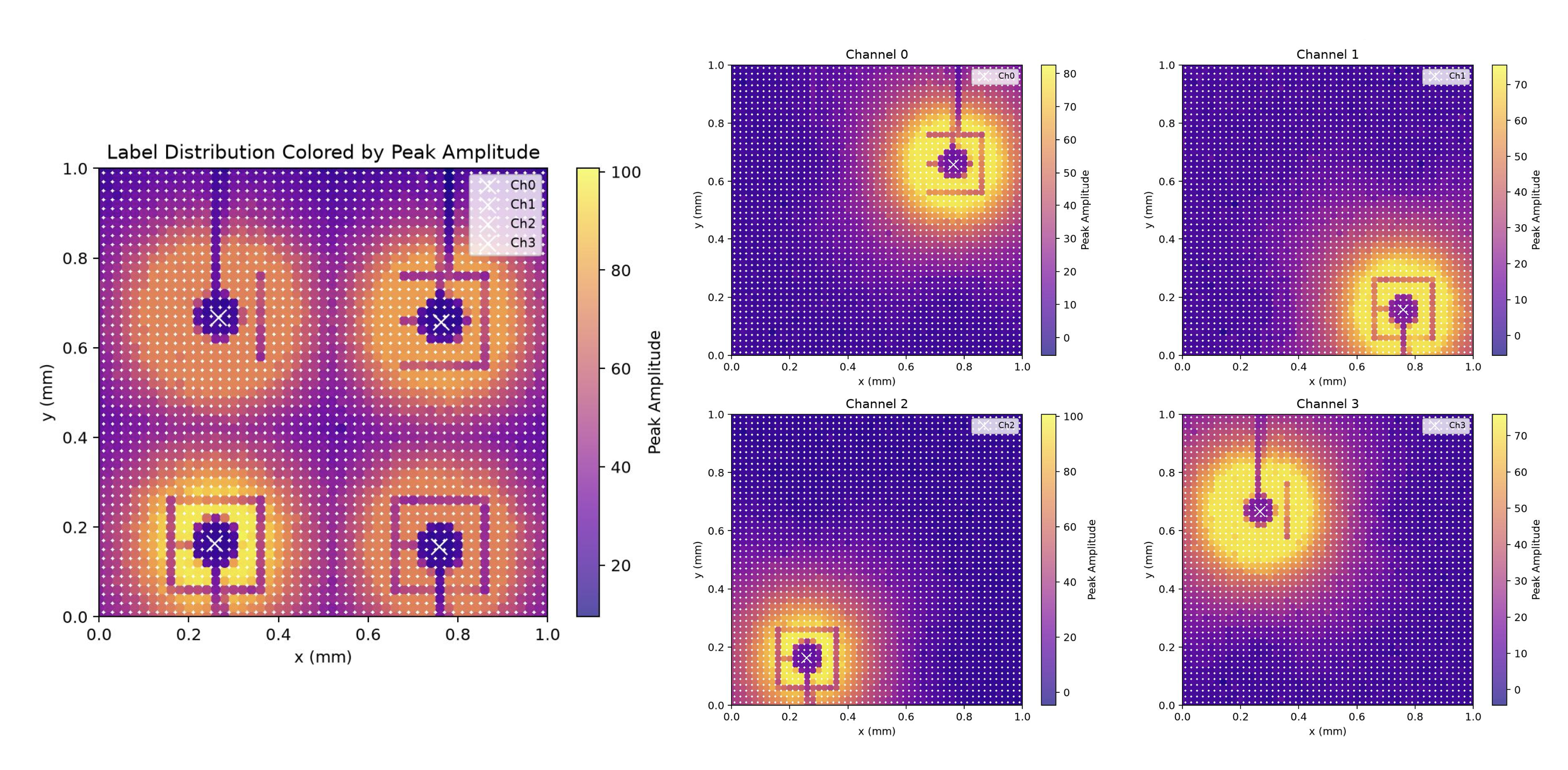}
    \caption{
        \textbf{Amplitude distribution for single-event waveforms at 190\volt.} Note the range of color bar in respective channel's subplot.
    }
    \label{fig:amplitude-dist}
\end{figure}

The attention bias injection through relative pad coordinate displacement assumes identical physical properties between pads. However, for the AC-LGAD involved in the TCT dataset, nuances between pad responses are observed. As shown in \autoref{fig:amplitude-dist}, the bottom-left channel (Channel 2) has stronger responses, while the top-left (Channel 3) and bottom-right (Channel 1) are the weakest.

\section{Automatic Computation of Pad Coordinate}
\begin{figure}[p]
    \centering
    \includegraphics[width=0.7\textwidth]{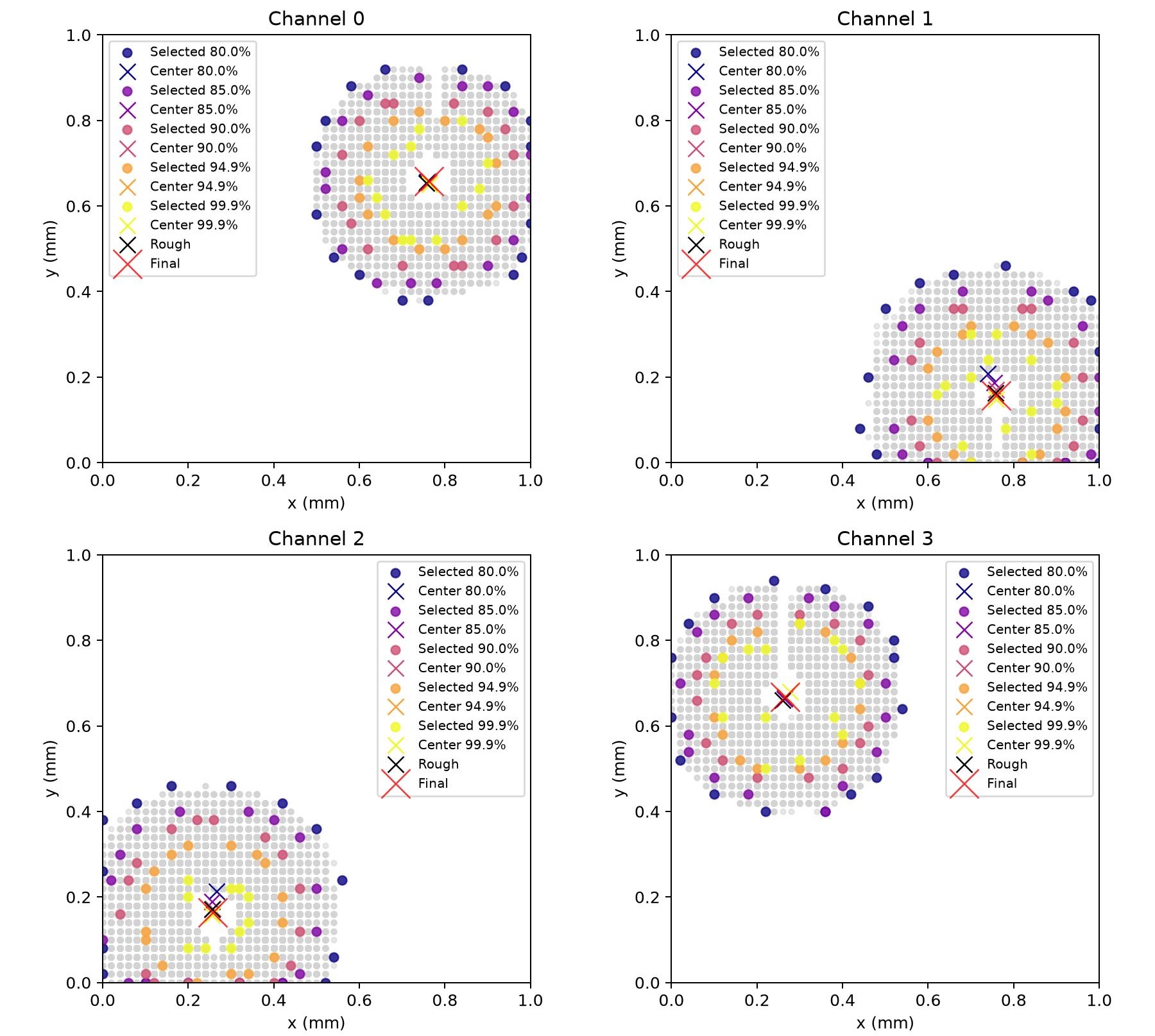}
    \caption{
        Pad coordinate estimation via circle fitting.
    }
    \label{fig:pad-estimate}
\end{figure}

GeoRead takes the physical coordinates of pads as an input. Assuming the waveform amplitude a pad receives is a function of distance from the pad center, we estimate the pad center by fitting circles on the convex hull of the distribution map of amplitudes above certain percentiles. The circles are expected to be approximately concentric. As shown in \autoref{fig:pad-estimate}, we perform fits for amplitudes above 80.0\%, 85.0\%, 90.0\%, 94.9\%, and 99.9\%. The two circles whose centers are furthest from the average are removed. The median of the remaining origins is considered the final estimate. Waveforms corresponding to the highest voltage (190\volt) are used for best SNR, and only train split data is used.

\section{More Results on Multi-event Prediction}
\begin{figure}[p]
    \centering
    \includegraphics[width=\textwidth]{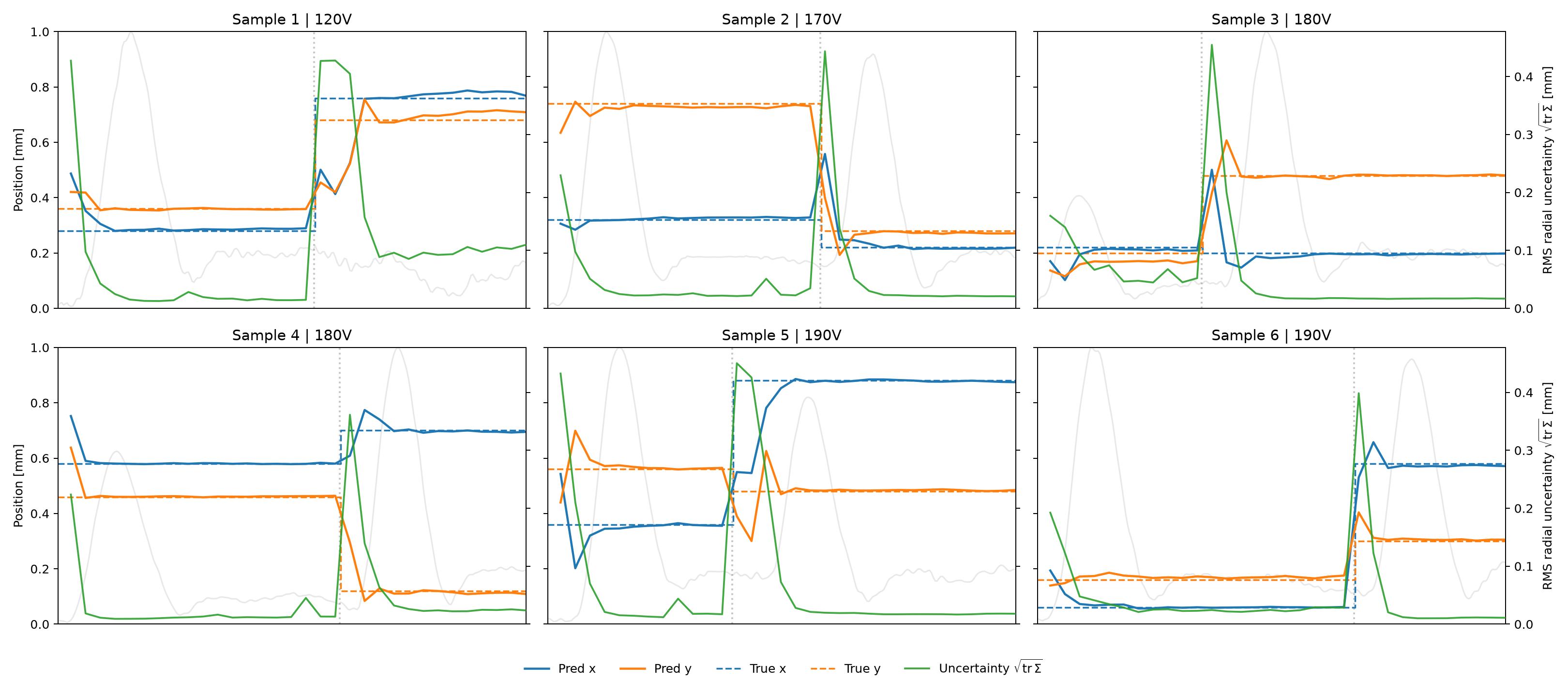}
    \caption{\textbf{Spatial-temporal predictions on mixed two-event sequences.} Each plot shows a superposed multi-event sample, with the waveform shape and model output at different time steps. The blue and orange curves show predicted x-y coordinates, and the dashed curves are ground truth coordinates. The green line represents the scalar uncertainty $\sqrt{\operatorname{tr}{\Sigma}}$. The uncertainty generally increases on low amplitude segments and reduces when more parts of a waveform are observed.
    All panels share the same axis limits.}
    \label{fig:multievent-more}
\end{figure}

We illustrate more qualitative results on multi-event predictions, as shown in 
\autoref{fig:multievent-more}.

\section{Model Predictions and Uncertainties over Scanned Region}
\begin{figure}[p]
    \centering
    \includegraphics[width=\textwidth]{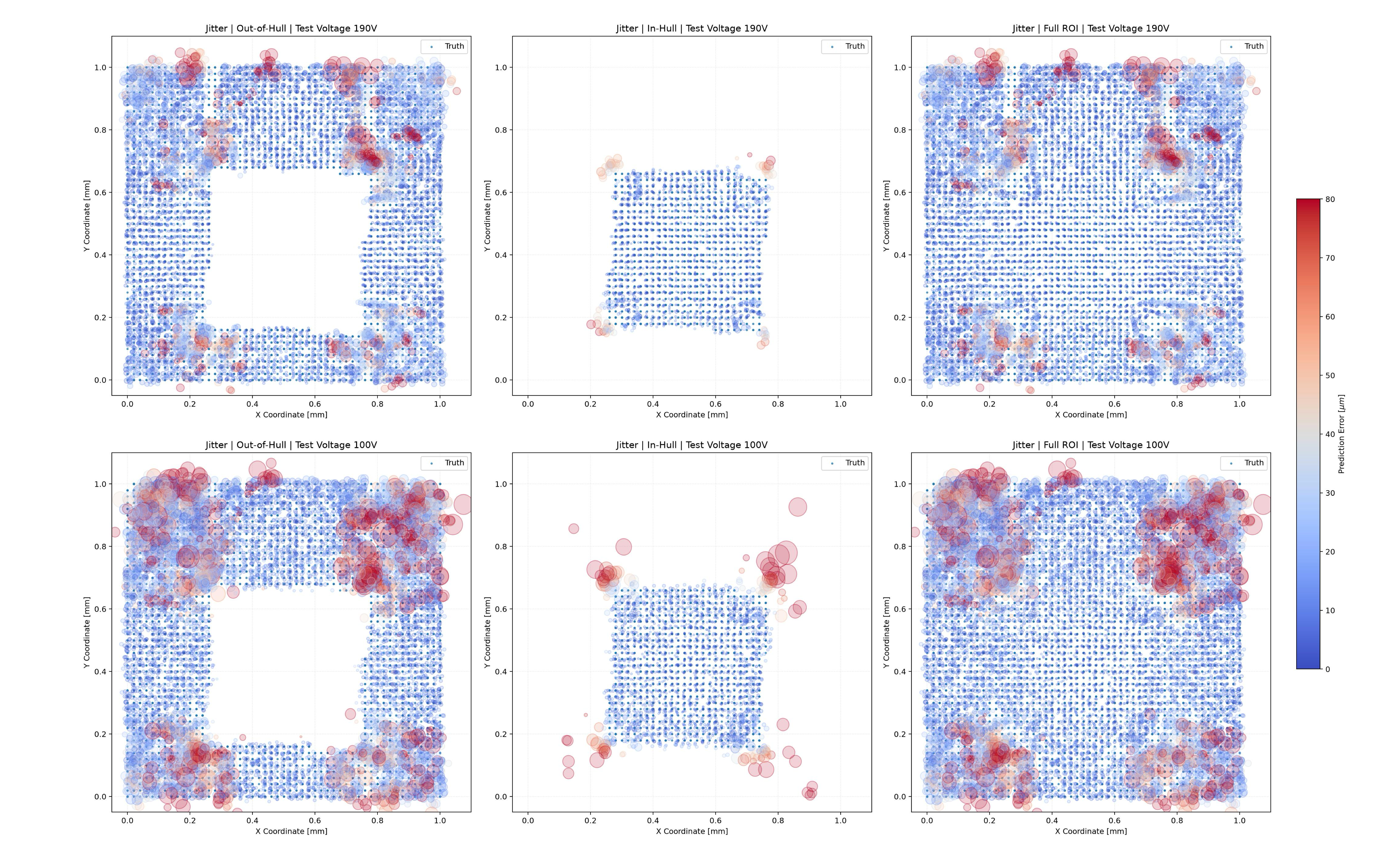}
    \caption{
        \textbf{Model's prediction for jittered single-event waveforms at various positions, at 190 and 100\volt.} Ground truth positions are indicated by solid blue dots. Each prediction is visualized by an ellipse, whose origin shows the predicted point and size shows the uncertainty. The ellipses are colored by reconstruction error, where blue ones indicate lower error. 
    }
    \label{fig:predictions}
\end{figure}

We visualize the model's predictions over the full scanned region using jittered single-event waveforms in \autoref{fig:predictions}. Larger reconstruction errors and predicted uncertainties are concentrated near the pad centers and readout wiring. In the pads,
the induced signal amplitudes are insensitive to
the hit position, providing limited spatial information for
reconstruction~\cite{TORNAGO2021165319}.

\section{Student-t and CPD Analysis on Single-event Spatial Resolution}
\begin{figure}[p]
    \centering
    \includegraphics[width=0.8\textwidth]{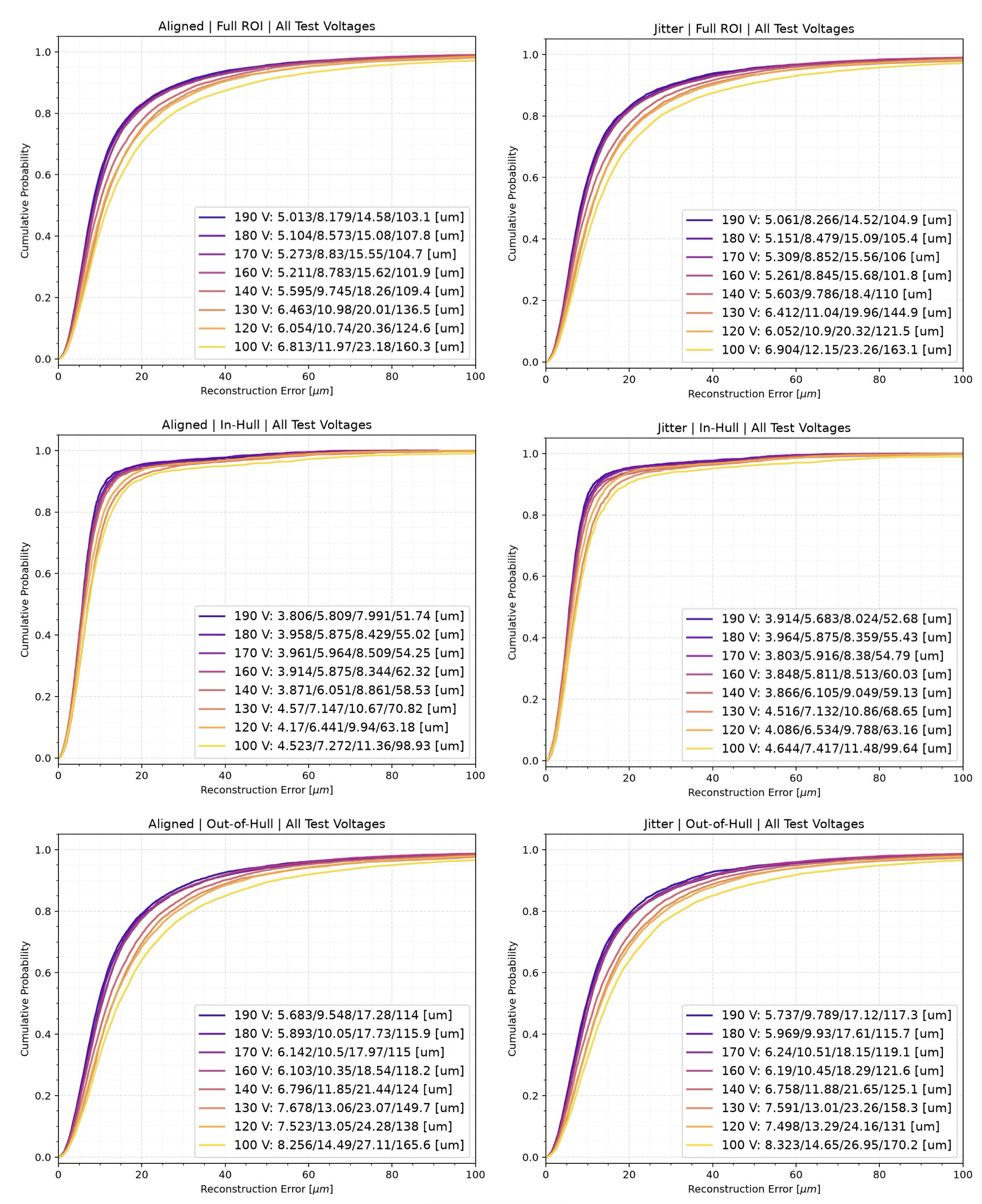}
    \caption{
        \textbf{Radial reconstruction residual as cumulative probability distribution (CPD) curves.} The x-axis represents radial residual, and the y-axis represents cumulative probability. The rows are full scanned region, in-hull region, and out-of-hull region, and the columns are aligned waveform test and jitter test, respectively. Each line in the legend represents a voltage point followed by 25th, 50th, 75th, and 99th percentile error in \mum. The closer the curve is to the upper-left corner, the better.
    }
    \label{fig:cpd-all}
\end{figure}

\begin{figure}[p]
    \centering
    \includegraphics[width=\textwidth]{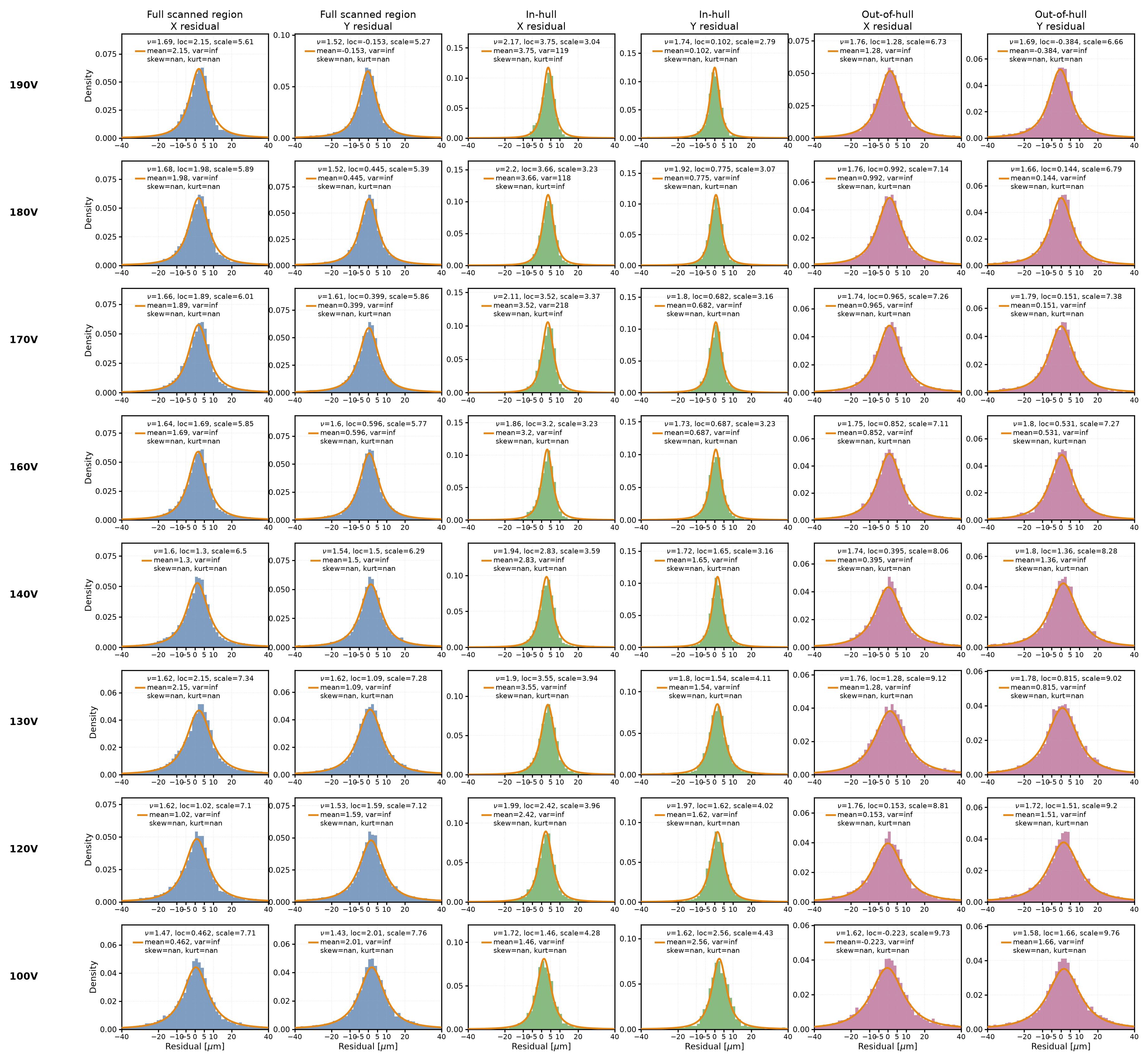}
    \caption{
        \textbf{Student-t fits for reconstruction residual across bias voltages for full scanned region, in-hull region, and out-of-hull region.} 
        The x-axis represents residual of model prediction in \mum, and the y-axis represents probability density. The test waveforms here are aligned, single-event ones.
    }
    \label{fig:student-t}
\end{figure}

\begin{figure}[p]
    \centering
    \includegraphics[width=\textwidth]{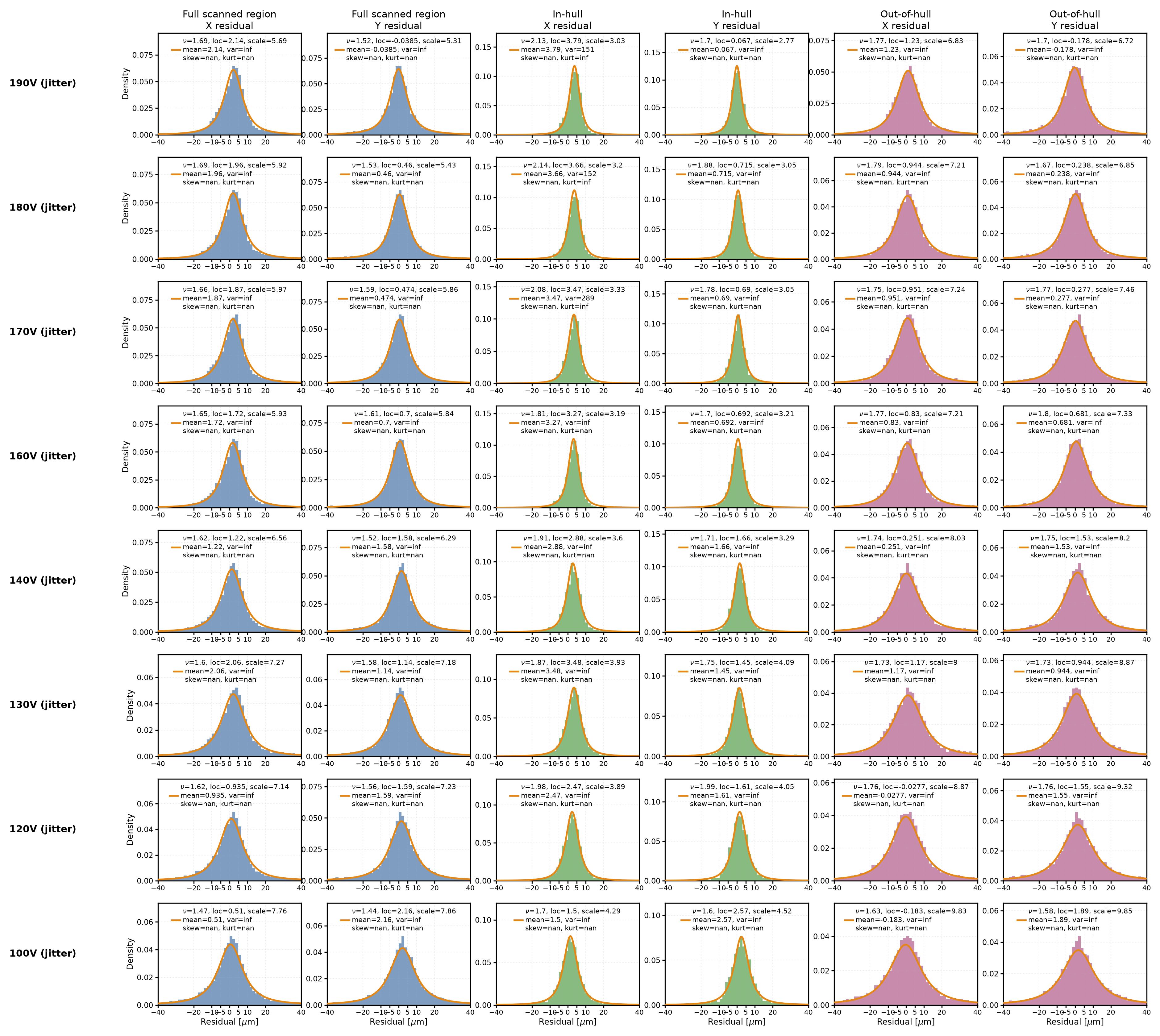}
    \caption{
        \textbf{Student-t fits for reconstruction residual evaluated on time-jitter waveforms, across bias voltages for full scanned region, in-hull region, and out-of-hull region.} 
        The x-axis represents residual of model prediction in \mum, and the y-axis represents probability density. The test waveforms here are jittered, single-event ones.
    }
    \label{fig:student-t-jittter}
\end{figure}

We perform Student-t analysis on single-event reconstruction residual distribution on full scanned region, in-hull region, and out-of-hull region. 
The results for aligned waveforms are shown in \autoref{fig:student-t}, and the results for jittered waveforms are shown in \autoref{fig:student-t-jittter}.
The radial residual distributions are also shown as cumulative probability distributions (CPD) in \autoref{fig:cpd-all}.


\clearpage
\bibliographystyle{JHEP}
\bibliography{merge.bib}

@Article{	  cartiglia2020,
  title		= {{LGAD} designs for Future PArticle Trackers},
  journal	= {Nuclear Instruments and Methods in Physics Research
		  Section A: Accelerators, Spectrometers, Detectors and
		  Associated Equipment},
  volume	= {979},
  pages		= {164383},
  year		= {2020},
  issn		= {0168-9002},
  doi		= {10.1016/j.nima.2020.164383},
  url		= {https://www.sciencedirect.com/science/Article/pii/S0168900220307804},
  author	= {N. Cartiglia and R. Arcidiacono and G. Borghi and M.
		  Boscardin and M. Costa and Z. Galloway and F. Fausti and M.
		  Ferrero and F. Ficorella and M. Mandurrino and S. Mazza and
		  E.J. Olave and G. Paternoster and F. Siviero and H. F-W.
		  Sadrozinski and V. Sola and A. Staiano and A. Seiden and M.
		  Tornago and Y. Zhao}
}

@Misc{		  barone2025,
  author	= {A. Aoki and G. Barone and B. Kilminster and D. Li and G.
		  Giacomini and H. Goyal and A. Macchiolo and S. Segner and
		  J. Tang and A. Tricoli},
  title		= {Studies on {ML} processing and compression of signal
		  shared {AC-LGADs}},
  year		= {2025},
  howpublished	= {\href{https://indico.cern.ch/event/1507215/contributions/6538685/}{3rd
		  DRD3 Week, CERN}}
}

@Misc{		  baronedrd3,
  author	= {G. Barone and others},
  title		= {From analog readout to {ML}-processed Silicon Device
		  signal-sharing and {LGADs} at {BNL}},
  year		= {2024},
  howpublished	= {\href{https://indico.cern.ch/event/1507215/contributions/6538685/}{1st
		  DRD3 Week on Solid State Detectors R$\&$D}}
}

@Misc{		  cms_tdr_mtd,
  author	= {{CMS Collaboration}},
  title		= {A {MIP} Timing Detector for the {CMS} Phase-2 Upgrade},
  institution	= {CERN},
  number	= {CERN-LHCC-2019-003, CMS-TDR-020},
  year		= {2019},
  howpublished	= {\href{https://cds.cern.ch/record/2667167}{CERN-LHCC-2019-003, CMS-TDR-020}}
}

@Article{	  cmscollaboration_2008,
  author	= {{CMS Collaboration}},
  title		= {The {CMS} experiment at the {CERN LHC}},
  journal	= {Journal of Instrumentation},
  volume	= {3},
  number	= {08},
  pages		= {S08004},
  year		= {2008},
  doi		= {10.1088/1748-0221/3/08/S08004},
  url		= {https://doi.org/10.1088/1748-0221/3/08/S08004}
}

@Article{	  ecfa_roadmap_2021,
  author	= {{ECFA Detector R$\&$D Roadmap Process Group}},
  title		= {The 2021 {ECFA} detector research and development
		  roadmap},
  journal	= "CERN-ESU-017",
  address	= "Geneva",
  year		= "2020",
  url		= "https://cds.cern.ch/record/2784893",
  doi		= "10.17181/CERN.XDPL.W2EX"
}

@Article{	  gautier2021165599,
  author	= {F. Gautier and A. Greeley and S. G. Kanekal and T. Isidori
		  and G. Legras and N. Minafra and A. Novikov and C. Royon
		  and Q. Schiller},
  title		= {A novel technique for real-time ion identification and
		  energy measurement for in situ space instrumentation},
  journal	= {Nuclear Instruments and Methods in Physics Research
		  Section A: Accelerators, Spectrometers, Detectors and
		  Associated Equipment},
  volume	= {1012},
  pages		= {165599},
  year		= {2021},
  doi		= {10.1016/j.nima.2021.165599},
  url		= {https://www.sciencedirect.com/science/article/pii/S0168900221005842}
}

@Article{	  heller2022_aclgad_bnl_hpk_jinst,
  doi		= {10.1088/1748-0221/17/05/P05001},
  url		= {https://doi.org/10.1088/1748-0221/17/05/P05001},
  year		= {2022},
  month		= {may},
  publisher	= {IOP Publishing},
  volume	= {17},
  number	= {05},
  pages		= {P05001},
  author	= {Heller, R. and Madrid, C. and Apresyan, A. and Brooks,
		  W.K. and Chen, W. and D'Amen, G. and Giacomini, G. and
		  Goya, I. and Hara, K. and Kita, S. and Los, S. and Molnar,
		  A. and Nakamura, K. and Pe{\~n}a, C. and San Mart{\'i}n, C.
		  and Tricoli, A. and Ueda, T. and Xie, S.},
  title		= {Characterization of {BNL} and {HPK AC-LGAD} sensors with a
		  120 {GeV} proton beam},
  journal	= {Journal of Instrumentation}
}

@Article{	  investigationchargesharing,
  author	= {Siviero, F. and Arcidiacono, R. and Cartiglia, N. and
		  Costa, M. and Ferrero, M. and Lanteri, L. and Madrid, C.
		  and Menzio, L. and Mulargia, R. and Sola, V.},
  title		= {Machine learning for precise hit position reconstruction
		  in Resistive Silicon Detectors},
  journal	= {Journal of Instrumentation},
  volume	= {19},
  number	= {01},
  pages		= {C01028},
  year		= {2024},
  doi		= {10.1088/1748-0221/19/01/C01028},
  url		= {https://doi.org/10.1088/1748-0221/19/01/C01028}
}

@Article{	  lgad1,
  author	= {G. Giacomini and W. Chen and F. Lanni and A. Tricoli},
  title		= {Development of a technology for the fabrication of
		  Low-Gain Avalanche Diodes at {BNL}},
  journal	= {Nuclear Instruments and Methods in Physics Research
		  Section A: Accelerators, Spectrometers, Detectors and
		  Associated Equipment},
  volume	= {934},
  pages		= {52--57},
  year		= {2019},
  doi		= {10.1016/j.nima.2019.04.073},
  url		= {https://doi.org/10.1016/j.nima.2019.04.073}
}

@Article{	  lgad2,
  author	= {G. Giacomini and W. Chen and G. D'Amen and E. Rossi and
		  A. Tricoli and A. Apresyan and R. Heller and R. Lipton and
		  K. Di Petrillo and C. Gee and S. Mazza and H. Sadrozinski
		  and B. Schumm and A. Seiden and Y. Zhao and R. Islam},
  journal	= {2021 IEEE Nuclear Science Symposium and Medical Imaging
		  Conference (NSS/MIC)},
  title		= {Fabrication of Different {LGAD}-Based Devices at {BNL}},
  year		= {2021},
  volume	= {},
  number	= {},
  pages		= {1-4},
  doi		= {10.1109/NSS/MIC44867.2021.987559}
}

@Article{	  mandurrino2020_nima959,
  title		= {Analysis and numerical design of Resistive {AC}-Coupled
		  Silicon Detectors {(RSD)} for {4D} particle tracking},
  journal	= {Nuclear Instruments and Methods in Physics Research
		  Section A: Accelerators, Spectrometers, Detectors and
		  Associated Equipment},
  volume	= {959},
  pages		= {163479},
  year		= {2020},
  issn		= {0168-9002},
  doi		= {10.1016/j.nima.2020.163479},
  url		= {https://www.sciencedirect.com/science/article/pii/S0168900220300851},
  author	= {M. Mandurrino and R. Arcidiacono and M. Boscardin and N.
		  Cartiglia and G.-F. {Dalla Betta} and M. Ferrero and F.
		  Ficorella and L. Pancheri and G. Paternoster and F. Siviero
		  and V. Sola and A. Staiano and A. Vignati}
}

@Article{	  pellegrini2014,
  author	= {G. Pellegrini and P. Fernandez-Martinez and M. Baselga and
		  D. Quirion and M. Ullan and J. M. Lopez and C. Fleta},
  title		= {Technology developments and first measurements of Low Gain
		  Avalanche Detectors {(LGAD)} for high energy physics
		  applications},
  journal	= {Nuclear Instruments and Methods in Physics Research
		  Section A: Accelerators, Spectrometers, Detectors and
		  Associated Equipment},
  volume	= {765},
  pages		= {12--16},
  year		= {2014},
  doi		= {10.1016/j.nima.2014.06.008},
  url		= {https://www.sciencedirect.com/science/article/pii/S0168900214007128}
}

@Article{	  royon2024168886,
  author	= {C. Royon and F. Gautier},
  title		= {Fast timing detectors with applications in cosmic ray
		  physics and medical science},
  journal	= {Nuclear Instruments and Methods in Physics Research
		  Section A: Accelerators, Spectrometers, Detectors and
		  Associated Equipment},
  volume	= {1058},
  pages		= {168886},
  year		= {2024},
  doi		= {10.1016/j.nima.2023.168886},
  url		= {https://www.sciencedirect.com/science/article/pii/S016890022300877X}
}

@Article{	  siviero2021,
  doi		= {10.1088/1748-0221/16/03/P03019},
  url		= {https://doi.org/10.1088/1748-0221/16/03/P03019},
  year		= {2021},
  month		= {mar},
  publisher	= {IOP Publishing},
  volume	= {16},
  number	= {03},
  pages		= {P03019},
  author	= {Siviero, F. and Arcidiacono, R. and Cartiglia, N. and
		  Costa, M. and Ferrero, M. and Legger, F. and Mandurrino, M.
		  and Sola, V. and Staiano, A. and Tornago, M.},
  title		= {First application of machine learning algorithms to the
		  position reconstruction in Resistive Silicon Detectors},
  journal	= {Journal of Instrumentation}
}

@Article{	  tornago2021_nima1003_165319,
  author	= {M. Tornago and R. Arcidiacono and N. Cartiglia and M.
		  Costa and M. Ferrero and M. Mandurrino and F. Siviero and
		  V. Sola and A. Staiano and A. Apresyan and K. {Di Petrillo}
		  and R. Heller and S. Los and G. Borghi and M. Boscardin and
		  G.-F. {Dalla Betta} and F. Ficorella and L. Pancheri and G.
		  Paternoster and H. Sadrozinski and A. Seiden},
  title		= {Resistive {AC}-Coupled Silicon Detectors {(RSD)}:
		  Principles of operation and first characterization},
  journal	= {Nuclear Instruments and Methods in Physics Research
		  Section A: Accelerators, Spectrometers, Detectors and
		  Associated Equipment},
  volume	= {1003},
  pages		= {165319},
  year		= {2021},
  doi		= {10.1016/j.nima.2021.165319},
  url		= {https://www.sciencedirect.com/science/article/pii/S016890022100303X}
}

@Article{	  tornago2023,
  title		= {Silicon sensors with resistive read-out: Machine Learning
		  techniques for ultimate spatial resolution},
  journal	= {Nuclear Instruments and Methods in Physics Research
		  Section A: Accelerators, Spectrometers, Detectors and
		  Associated Equipment},
  volume	= {1047},
  pages		= {167816},
  year		= {2023},
  issn		= {0168-9002},
  doi		= {10.1016/j.nima.2022.167816},
  url		= {https://www.sciencedirect.com/science/article/pii/S0168900222011081},
  author	= {M. Tornago and F. Giobergia and L. Menzio and F. Siviero
		  and R. Arcidiacono and N. Cartiglia and M. Costa and M.
		  Ferrero and G. Gioachin and M. Mandurrino and V. Sola}
}

@Article{	  wu20192672,
  author	= {X. Wu and G. Ambrosi and P. Azzarello and B. Bergmann and
		  B. Bertucci and F. Cadoux and M. Campbell and M. Duranti
		  and M. Ionica and M. Kole and S. Krucker and G. Maehlum and
		  D. Meier and M. Paniccia and L. Pinsky and C. Plainaki and
		  S. Pospisil and T. Stein and P. A. Thonet and A. Tykhonov},
  title		= {Penetrating particle ANalyzer {(PAN)}},
  journal	= {Advances in Space Research},
  volume	= {63},
  number	= {8},
  pages		= {2672--2682},
  year		= {2019},
  doi		= {10.1016/j.asr.2019.01.012},
  url		= {https://doi.org/10.1016/j.asr.2019.01.012}
}

@Article{	  wu_2020,
  author	= {K. Wu and M. Zhao and T. Yang and J. G. da Costa and Z.
		  Liang and X. Shi},
  title		= {Design and fabrication of Low Gain Avalanche Detectors
		  {(LGAD)}: a {TCAD} simulation study},
  journal	= {Journal of Instrumentation},
  volume	= {15},
  number	= {03},
  pages		= {C03008},
  year		= {2020},
  doi		= {10.1088/1748-0221/15/03/C03008},
  url		= {https://doi.org/10.1088/1748-0221/15/03/C03008}
}

@article{TORNAGO2021165319,
title = {Resistive AC-Coupled Silicon Detectors: Principles of operation and first results from a combined analysis of beam test and laser data},
journal = {Nuclear Instruments and Methods in Physics Research Section A: Accelerators, Spectrometers, Detectors and Associated Equipment},
volume = {1003},
pages = {165319},
year = {2021},
issn = {0168-9002},
doi = {https://doi.org/10.1016/j.nima.2021.165319},
url = {https://www.sciencedirect.com/science/article/pii/S016890022100303X},
author = {M. Tornago and R. Arcidiacono and N. Cartiglia and M. Costa and M. Ferrero and M. Mandurrino and F. Siviero and V. Sola and A. Staiano and A. Apresyan and K. {Di Petrillo} and R. Heller and S. Los and G. Borghi and M. Boscardin and G.-F. {Dalla Betta} and F. Ficorella and L. Pancheri and G. Paternoster and H. Sadrozinski and A. Seiden}
}

@inproceedings{vaswani2017attention,
  title     = {Attention Is All You Need},
  author    = {Vaswani, Ashish and Shazeer, Noam and Parmar, Niki and Uszkoreit, Jakob and Jones, Llion and Gomez, Aidan N. and Kaiser, Lukasz and Polosukhin, Illia},
  booktitle = {Advances in Neural Information Processing Systems},
  volume    = {30},
  year      = {2017}
}

@article{SU2024127063,
title = {RoFormer: {Enhanced Transformer} with {Rotary Position Embedding}},
journal = {Neurocomputing},
volume = {568},
pages = {127063},
year = {2024},
issn = {0925-2312},
doi = {https://doi.org/10.1016/j.neucom.2023.127063},
url = {https://www.sciencedirect.com/science/article/pii/S0925231223011864},
author = {Jianlin Su and Murtadha Ahmed and Yu Lu and Shengfeng Pan and Wen Bo and Yunfeng Liu},
}

@InProceedings{qu2024particletransformerjettagging,
  title = 	 {Particle {Transformer} for {Jet Tagging}},
  author =       {Qu, Huilin and Li, Congqiao and Qian, Sitian},
  booktitle = 	 {Proceedings of the 39th International Conference on Machine Learning},
  pages = 	 {18281--18292},
  year = 	 {2022},
  editor = 	 {Chaudhuri, Kamalika and Jegelka, Stefanie and Song, Le and Szepesvari, Csaba and Niu, Gang and Sabato, Sivan},
  volume = 	 {162},
  series = 	 {Proceedings of Machine Learning Research},
  month = 	 {17--23 Jul},
  publisher =    {PMLR},
  url = 	 {https://proceedings.mlr.press/v162/qu22b.html},
}

@article{siviero2022rsd16ch,
  author        = {Siviero, F. and Giobergia, F. and Menzio, L. and Miserocchi, F. and Tornago, M. and Arcidiacono, R. and Cartiglia, N. and Costa, M. and Ferrero, M. and Gioachin, G. and Mandurrino, M. and Sola, V.},
  title         = {First experimental results of the spatial resolution of {RSD} pad arrays read out with a 16-ch board},
  journal       = {Nuclear Instruments and Methods in Physics Research Section A: Accelerators, Spectrometers, Detectors and Associated Equipment},
  volume        = {1041},
  pages         = {167313},
  year          = {2022},
  doi           = {10.1016/j.nima.2022.167313},
  eprint        = {2204.06388},
  archivePrefix = {arXiv},
  primaryClass  = {physics.ins-det},
  url           = {https://arxiv.org/abs/2204.06388}
}

@article{aoki2026resistive,
  author        = {Aoki, Alexander and Barone, Gaetano and Diehl, Leena and Giacomini, Gabriele and Gkougkousis, Vagelis and Goyal, Hanshal and Kher, Rohan and Li, Daniel and Macchiolo, Anna and Padnuik, Yevhenii and Senina, Daria and Sunnarborg, Samantha and Tang, Jessica and Tricoli, Alessandro and Wang, Lixing and Wong, Don C.},
  title         = {Machine Learning-Based Reconstruction for Resistive Silicon Sensors},
  journal       = {Journal of Instrumentation},
  volume        = {21},
  number        = {08},
  pages         = {C08015},
  year          = {2026},
  doi           = {10.1088/1748-0221/21/08/C08015},
  eprint        = {2607.11585},
  archivePrefix = {arXiv},
  primaryClass  = {hep-ex},
  url           = {https://arxiv.org/abs/2607.11585}
}

@article{source_chinaxiv_202602_00195,
  author = {Luo, Jia-Qi and Li, Han and Yue, Tian-Shuo and Zhang, De and Ma, Kuo and Fu, Changbo and He, Wan-Bing and Tang, Zebo and Liang, Hao and Liu, Yanwen and Ma, Yu-Gang},
  title = {End-to-end deep learning for high-precision
           {2D} hit reconstruction in {AC-LGAD} sensors},
  journal = {ChinaXiv preprint},
  note = {ChinaXiv:202602.00195},
  year={2026},
  doi = {10.12074/202602.00195},
  url = {https://chinaxiv.org/abs/202602.00195},
}

\end{document}